\documentclass[%
 reprint,
superscriptaddress,
 amsmath,amssymb,
 aps,
]{revtex4-2}

\usepackage{graphicx}
\usepackage{dcolumn}
\usepackage{bm}
\usepackage[mathlines]{lineno}
\usepackage{xspace}
\usepackage{bbding}
\usepackage{siunitx}
\usepackage{tikz,xcolor,hyperref}

\definecolor{lime}{HTML}{A6CE39}
\DeclareRobustCommand{\orcidicon}{
	\begin{tikzpicture}
	\draw[lime, fill=lime] (0,0) 
	circle [radius=0.16] 
	node[white] {{\fontfamily{qag}\selectfont \tiny ID}};
	\draw[white, fill=white] (-0.0625,0.095) 
	circle [radius=0.007];
	\end{tikzpicture}
	\hspace{-2mm}
}
\foreach \x in {A, ..., Z}{\expandafter\xdef\csname orcid\x\endcsname{\noexpand\href{https://orcid.org/\csname orcidauthor\x\endcsname}
			{\noexpand\orcidicon}}
}

\newcommand{\CP}{\ensuremath{CP}\xspace}
\newcommand{\strongCP}{strong-\ensuremath{CP}\xspace}
\newcommand{\runOneClimitMass}{\ensuremath{3.3\text{--}4.2~\si\micro {\rm eV}}\xspace}
\newcommand{\previouslimitMass}{\ensuremath{2.7\text{--}3.3~\si\micro {\rm eV}}\xspace}
\newcommand{\Tsys}{\ensuremath{T_{\rm sys}}\xspace}
\newcommand{\TsysEps}{\ensuremath{T_{\rm sys}/\epsilon}\xspace}
\newcommand{\Thot}{\ensuremath{T_{\rm hot}}\xspace}
\newcommand{\ToffEp}{\ensuremath{T_{\rm off}/\epsilon}\xspace}
\newcommand{\Paxion}{\ensuremath{P_{\rm axion}}\xspace}
\newcommand{\kboltz}{\ensuremath{k_{\rm B}}\xspace}
\newcommand{\gagamgam}{\ensuremath{g_{a\gamma\gamma}}\xspace}
\newcommand{\ggam}{\ensuremath{g_{\gamma}}\xspace}

\begin{document}

\preprint{APS/123-QED}

\title{Search for ``Invisible" Axion Dark Matter in the \runOneClimitMass Mass Range}
\author{C. Bartram}
\author{T. Braine}
\author{E. Burns}
\author{R. Cervantes}
\author{N. Crisosto}
\author{N. Du}
\author{H. Korandla}
\author{G. Leum}
\author{P. Mohapatra}
\author{T. Nitta}\email[Correspondence to: ]{tnitta1@uw.edu}
\author{L. J Rosenberg}
\author{G. Rybka}
\author{J.~Yang}%
  \affiliation{University of Washington, Seattle, Washington 98195, USA}

\author{John Clarke}
\author{I. Siddiqi}
  \affiliation{University of California, Berkeley, California 94720, USA}

\author{A. Agrawal}
\author{A. V. Dixit}
\affiliation{University of Chicago, Illinois 60637, USA}

\author{M. H. Awida} 
\author{A. S. Chou} 
\author{M. Hollister}
  \affiliation{Fermi National Accelerator Laboratory, Batavia, Illinois 60510, USA}

\author{S. Knirck}
\author{A. Sonnenschein} 
\author{W. Wester} 
  \affiliation{Fermi National Accelerator Laboratory, Batavia, Illinois 60510, USA}
%
\author{J.~R.~Gleason}
\author{A. T. Hipp}
\author{S. Jois}
\author{P. Sikivie}
\author{N. S. Sullivan}
\author{D. B. Tanner}
  \affiliation{University of Florida, Gainesville, Florida 32611, USA}

\author{E. Lentz}
  \affiliation{University of G\"{o}ttingen, G\"{o}ttingen 37077, Germany}

\author{R. Khatiwada}
  \affiliation{Illinois Institute of Technology, Chicago, Illinois 60616, USA}
  \affiliation{Fermi National Accelerator Laboratory, Batavia, Illinois 60510, USA}

\author{G. Carosi}
\author{N. Robertson}
\author{N. Woollett}
  \affiliation{Lawrence Livermore National Laboratory, Livermore, California 94550, USA}

\author{L. D. Duffy}
  \affiliation{Los Alamos National Laboratory, Los Alamos, New Mexico 87545, USA}

\author{C. Boutan}
\author{M. Jones}
\author{B. H. LaRoque}
\author{N. S.~Oblath}
\author{M. S. Taubman}
  \affiliation{Pacific Northwest National Laboratory, Richland, Washington 99354, USA}

\author{E. J. Daw}
\author{M. G. Perry}
  \affiliation{University of Sheffield, Sheffield S3 7RH, UK}
  
\author{J. H. Buckley}
\author{C. Gaikwad}
\author{J. Hoffman}
\author{K. W. Murch}
  \affiliation{Washington University, St. Louis, Missouri 63130, USA}

\author{M. Goryachev}
\author{B. T. McAllister}
\author{A. Quiskamp}
\author{C. Thomson}
\author{M. E. Tobar}

  \affiliation{University of Western Australia, Perth, Western Australia 6009, Australia}

\collaboration{ADMX Collaboration}\noaffiliation


\begin{abstract}
We report the results from a haloscope search for axion dark matter in the \runOneClimitMass mass range. This search excludes the axion-photon coupling predicted by one of the benchmark models of ``invisible" axion dark matter, the KSVZ model. This sensitivity is achieved using a large-volume cavity, a superconducting magnet, an ultra low noise Josephson parametric amplifier, and sub-Kelvin temperatures. The validity of our detection procedure is ensured by injecting and detecting blind synthetic axion signals. 
\end{abstract}

\maketitle

In the Standard Model of particle physics, the amount of Charge-Parity (\CP) violation by the strong 
interactions is set by an angle $\theta$ whose value is expected to be of order one~\cite{RevModPhys.93.015004}. However, the upper limit on the neutron electric dipole moment~\cite{PhysRevLett.124.081803} requires $\theta <  5 \times 10^{-11}$.  This discrepancy is called the \strongCP problem. The existence of a new global axial U(1) symmetry, proposed by Peccei and Quinn (PQ) ~\cite{Peccei1977June}, would solve the \strongCP problem. This symmetry must be spontaneously broken, implying the existence of a pseudo Nambu-Goldstone boson, called the axion~\cite{Peccei1977June,Weinberg:1977ma,Wilczek:1977pj}. The relic axions produced during the QCD phase transition in the early universe~\cite{PRESKILL1983127,ABBOTT1983133,DINE1983137} satisfy all the requirements of dark matter~\cite{PhysRevLett.50.925}. Thus the hypothetical axion solves both the \strongCP and dark matter problems. In the scenario in which the PQ symmetry breaks before cosmological inflation (pre-inflationary scenario), the relic axion abundance is determined only by the initial amplitude ($\theta_0$) and mass of the axion field. The abundance of dark matter in the Lambda cold dark matter model is naturally explained by an axion mass above $\sim 0.1~\si\micro$eV for $\theta_0>0.1$~\cite{Borsanyi2016,PhysRevD.96.095001}. In the post-inflationary scenario, where the PQ symmetry breaks after cosmological inflation, most calculations suggest that the axion mass lies in the $\mathcal{O}(1\text{--}100)~\si\micro$eV range~\cite{PhysRevD.83.123531,PhysRevD.91.065014,PhysRevD.92.034507,Fleury_2016,Bonati2016,PETRECZKY2016498,Borsanyi2016,PhysRevLett.118.071802,Klaer_2017,PhysRevD.96.095001,PhysRevLett.124.161103,10.21468/SciPostPhys.10.2.050,buschmann2021dark}. The axion has a coupling to two photons, and the numerical values are represented by two benchmark models, the Kim-Shifman-Vainshtein-Zakharov (KSVZ) \cite{Kim:1979if,Shifman:1979if} and Dine-Fischler-Srednicki-Zhitnitsky (DFSZ) \cite{Dine:1981rt,Zhitnitsky:1980tq} models. Since those axion-photon couplings are expected to be small, $\mathcal{O}(10^{-17}\text{--}10^{-12})~{\rm GeV^{-1}}$, axions predicted by the models are called ``invisible" axions~\cite{PhysRevLett.51.1415}.

To date, only the Axion Dark Matter eXperiment (ADMX) \cite{PhysRevD.64.092003,PhysRevD.69.011101,Asztalos_2002,PhysRevLett.104.041301,PhysRevLett.120.151301,PhysRevLett.124.101303} has attained a sensitivity to the DFSZ model, which is a particularly well-motivated model because it can be grand unified. ADMX is a haloscope experiment \cite{PhysRevLett.51.1415,RevModPhys.82.557,RevModPhys.93.015004} searching for axions within the local halo with a cold resonant cavity immersed in a static magnetic field. Maxwell's equations modified to include the axion-photon interaction imply that an oscillating axion field ($\phi$) in a static magnetic field ($\vec{B}$) induces an oscillating electric current, $\vec{j}_a = \gagamgam\vec{B}\partial_t \phi$, where \gagamgam is the coupling of the axion to two photons. The electric current $\vec{j}_a$ oscillates with a frequency $E/h$, where $E$ is the sum of the mass ($m$) and kinetic energy of dark matter axions and $h$ is the Planck's constant. $E/h\approx mc^2/h$ because halo axions are non-relativistic. The induced currents resonantly drive electromagnetic modes of the cavity with a resonant frequency equal to the frequency $E/h$ of axion field oscillations. This signal is extracted by an antenna, amplified by several amplifiers, and sampled by a digitizer. Because the power from the axion signal is extremely small due to the minuscule axion-photon coupling, physical temperatures and electronic noise from the amplifiers need to be as low as possible. 

Previous reports by the ADMX collaboration have excluded masses over \previouslimitMass for the DFSZ model~\cite{PhysRevLett.120.151301,PhysRevLett.124.101303}. This Letter reports results of the search for axions in the \runOneClimitMass mass range. 

\begin{figure}[phtb]
    \centering
    \includegraphics[width=\linewidth]{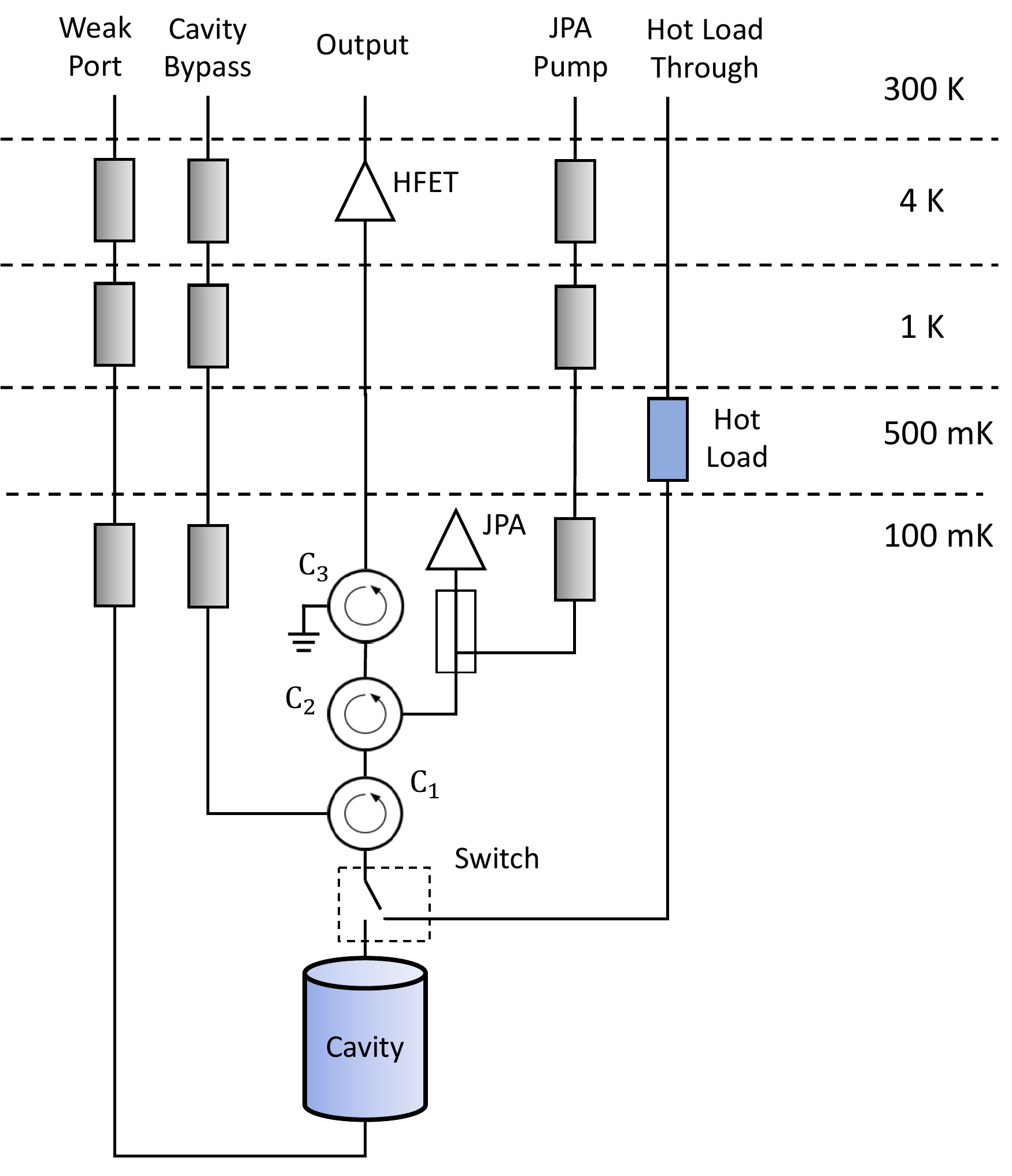}
    \caption{The ADMX RF diagram. $C_1$, $C_2$, and $C_3$ are circulators. The JPA is connected with a pump power line via a directional coupler. The gray-colored rectangular boxes denote cryogenic attenuators. The switch is connected to the cavity during data-taking, the bypass for the hot-load is used for system noise calibrations.}
    \label{fig:rf_layout}
\end{figure}

The ADMX experimental apparatus consists of a 136~$\ell$ cylindrical copper-plated stainless  steel microwave cavity in a 7.5~T superconducting magnet. Two movable bulk copper rods inside the cavity tune its resonant frequency. A variable depth antenna at the top of the cavity picks up RF signals inside the cavity. The coupling to the cavity is kept in a critically- or over-coupled state by varying the insertion depth of the antenna to maximize the sensitivity to axion signals. A simplified RF diagram for the ADMX apparatus is shown in Fig.~\ref{fig:rf_layout}. The RF signals extracted from the cavity pass through two circulators and are amplified by the first stage amplifier, a Josephson Parametric Amplifier (JPA)~\cite{Siddiqi_2004}. The JPA achieves parametric amplification using the four-wave mixing produced by the non-linearity of its SQUID (Superconducting QUantum Interference Device) loops. The JPA exhibits ultra-low noise performance, just above the quantum limit, by adding noise only from the thermal population of the mode at the idler frequency.~\cite{9134828}. The JPA is operated in a phase-preserving mode with a static current run through a nearby flux loop to bias the SQUIDs and a pump tone offset by 300~kHz from the resonant frequency of the cavity. Because the circulators and the JPA are sensitive to external magnetic fields, they are placed in a magnetic-field-free region generated by a bucking coil magnet designed to cancel the stray magnetic field from the main superconducting magnet. Empirically, we found that for a fixed bucking coil current, variations smaller than 0.2~A in the main magnet produced no significant changes in the JPA performance. Additionally, the JPA is inside a three-layer $\mu$-metal shield to attenuate any remaining fields. By fine-tuning the bias current and pump power, we achieved a power gain of $15\text{--}30$ dB across the frequency range. Amplified RF signals propagate through two circulators and are further amplified by the second stage amplifier, a Heterostructure Field Effect Transistor (HFET) amplifier, model number $\rm LNF\text{--}LNC0.6\_2A$ \cite{LNFLNC026A}, placed at the 4~K stage. At room temperature, these RF signals are amplified by a HFET amplifier, mixed down to 10.7~MHz, and sampled by the digitizer. The resonant modes of the cavity and the antenna coupling are monitored by a vector network analyzer (VNA) via the weak port and cavity bypass RF lines. A dilution refrigerator maintained an approximate temperature of 110~mK at the mixing chamber, enabling temperatures of 150~mK at the cavity and 120~mK at the JPA and circulators.  Further details can be found in Ref.~\cite{khatiwada2020axion}.

The power from axion conversion inside the cavity~\cite{PhysRevLett.51.1415} is
\begin{equation}
\begin{split}
    \Paxion=7.7\times 10^{-23}~\text{W} \left(\frac{V}{136~\ell}\right)\left(\frac{B}{7.5~\text{T}}\right)^2\left(\frac{C}{0.4}\right)\\
    \times\left(\frac{\ggam}{0.36}\right)^2\left(\frac{\rho_a}{0.45~\text{GeV/cc}}\right)\left(\frac{f}{1~\text{GHz}}\right)\left(\frac{Q_L}{80,000}\right).
\end{split}
\label{Eq:Axion_Power}
\end{equation}
Here, $V$ is the volume of the cavity, $B$ is the magnitude of the magnetic field, $C$ is the form factor representing the overlap between the cavity resonant mode and the magnetic field, \ggam is the model-dependent numerical constant $-0.97$ (0.36) for the KSVZ (DFSZ) model which determines, along with the axion decay constant $f_a$, the axion coupling to two photons $\gagamgam=\alpha \ggam/\pi f_a$, $\rho_a$ is the expected dark matter density in the cavity, $f$ is the frequency of the photon induced by the axion field, and $Q_L$ is the loaded quality factor of the cavity.

The signal-to-noise ratio (SNR) is used as a metric of the sensitivity of the experiment \cite{doi:10.1063/1.1770483}:
\begin{equation}
{\rm SNR} = \frac{\Paxion}{\kboltz\TsysEps}\sqrt{\frac{t}{b}},
\end{equation}
where \kboltz is the Boltzmann constant, \Tsys is the system noise temperature, which is defined as the sum of the physical and electronic noise temperatures, $\epsilon$ is the transmission efficiency between the cavity and the JPA, $t$ is the integration time, and $b$ is the detection bandwidth $b = f/Q_a \sim f/10^6$, the expected signal energy spread of non-relativistic axion dark matter with a velocity of $\sim 10^{-3}c$.

\Tsys is measured with the SNR Improvement (SNRI) method. The SNRI is given by 
\begin{equation}
{\rm SNRI} = \Tsys^{\rm off}/\Tsys^{\rm on} = \frac{G_{\rm on}}{G_{\rm off}}\frac{P_{\rm off}}{P_{\rm on}},
\end{equation}
where $G_{\rm on (off)}$ and $P_{\rm on (off)}$ are the total gain of the RF chain measured by the VNA and the power spectral density of the RF chain measured by the digitizer for the JPA in the on (off) state, respectively. As an overcoupled resonator, the JPA acts as a lossless mirror for the signal when the JPA is off. The SNRI was typically 7.5 dB. $\Tsys^{\rm off}$ is measured by the $y$-factor method~\cite{wilson2011techniques} utilizing a noise source placed at the 500~mK stage (labeled as the ``Hot Load''). The hot load temperature, \Thot, can be varied between $0.5\text{--}4$ K. During a $y$-factor measurement, the input of the cold receiver is connected to the hot load by flipping a switch in the receiver so that thermal photon from the hot load is detected. There is a linear relation between \Thot and the digitized power when the JPA is off:
\begin{equation}
P = G_{\rm off}b\kboltz (\Thot\epsilon_h+T_{\rm sys}^{\rm off}).
\end{equation}
Here, $\epsilon_h$ is the total transmission efficiency between the hot load and the JPA. Hence, one can obtain \ToffEp from the $y$ intersection of a linear fit. We assume $\epsilon_h $ is equal to $\epsilon$ since $\epsilon_h$ is dominated by losses in the circulators. During the data-taking period, \ToffEp was measured by the $y$-factor method every few months. The results were stable over time around 3.5~K, though they varied with frequency by 0.2~K over the frequency range. From the above, the typical \TsysEps when the JPA is on was calculated to be 600~mK.

The data described here were acquired between October 2019 and May 2021. We aimed to probe axions with one (two) times DFSZ coupling between 950 and 1020 (800 and 950)~MHz. Throughout the operation, Synthetic Axion Generated signals (SAGs), which are created using lower power RF tones, were injected into the cavity via its weak port to ensure the robustness of the experiment. Two types of SAGs were injected: calibration SAGs, which were intended to verify the integrity of the receiver chain and analysis framework, and blinded SAGs to practice the full candidate evaluation procedure. 

The explored frequency range was divided into 14 ``nibbles,'' narrow frequency ranges, typically 10~MHz in width. The specific procedure repeated for each nibble is as follows. First, digitizations for the entire nibble frequency range were performed by moving both tuning rods symmetrically. The integration time was 100 seconds. The spectral width was 50~kHz for each scan, and the data were averaged with 100~Hz bin resolution. Next, an analysis was performed to check for axion-like excesses (candidates) above the noise. 

The shape of the detected background is primarily determined by two factors: the shape generated by the room temperature receiver and the JPA standing wave due to the imperfect isolation of circulators, $C_1$ and $C_2$. The former is time independent and was removed by a reference shape measured at the beginning of data-taking. The latter varies when the bias current of the JPA is changed and was removed by a six-order Pad\'{e}-approximant performed for each spectrum. The flattened spectra were scaled by the estimated \TsysEps to obtain the correct power scale and were convolved with the expected axion shape to improve the sensitivity to axions. The spectra were co-added into a ``grand spectrum" to make use of all recorded spectra. Typically, $\mathcal{O}(10)$ candidates are found within a nibble because of statistical fluctuations and calibration SAGs. The criteria to select candidates are described in Ref.~\cite{PhysRevD.103.032002}. Accordingly, the candidates were scanned further, a rescan, to check whether they are persistent. The rescan data are later included in the analysis, and a Monte Carlo study indicates that this procedure may induce a small bias of less than 3\% on the resulting extracted axion-photon coupling limits in specific circumstances. After the first rescan, the calibration SAGs were turned off and the second rescan was performed to confirm whether the candidates were SAGs or true signals. If all SAGs were identified and there was no candidate left, the data-taking moved on to the next nibble. However, if there still remained one or more candidates, more rigorous tests were performed as described below. 

\begin{figure}[phtb]
    \centering
    \includegraphics[width=\linewidth]{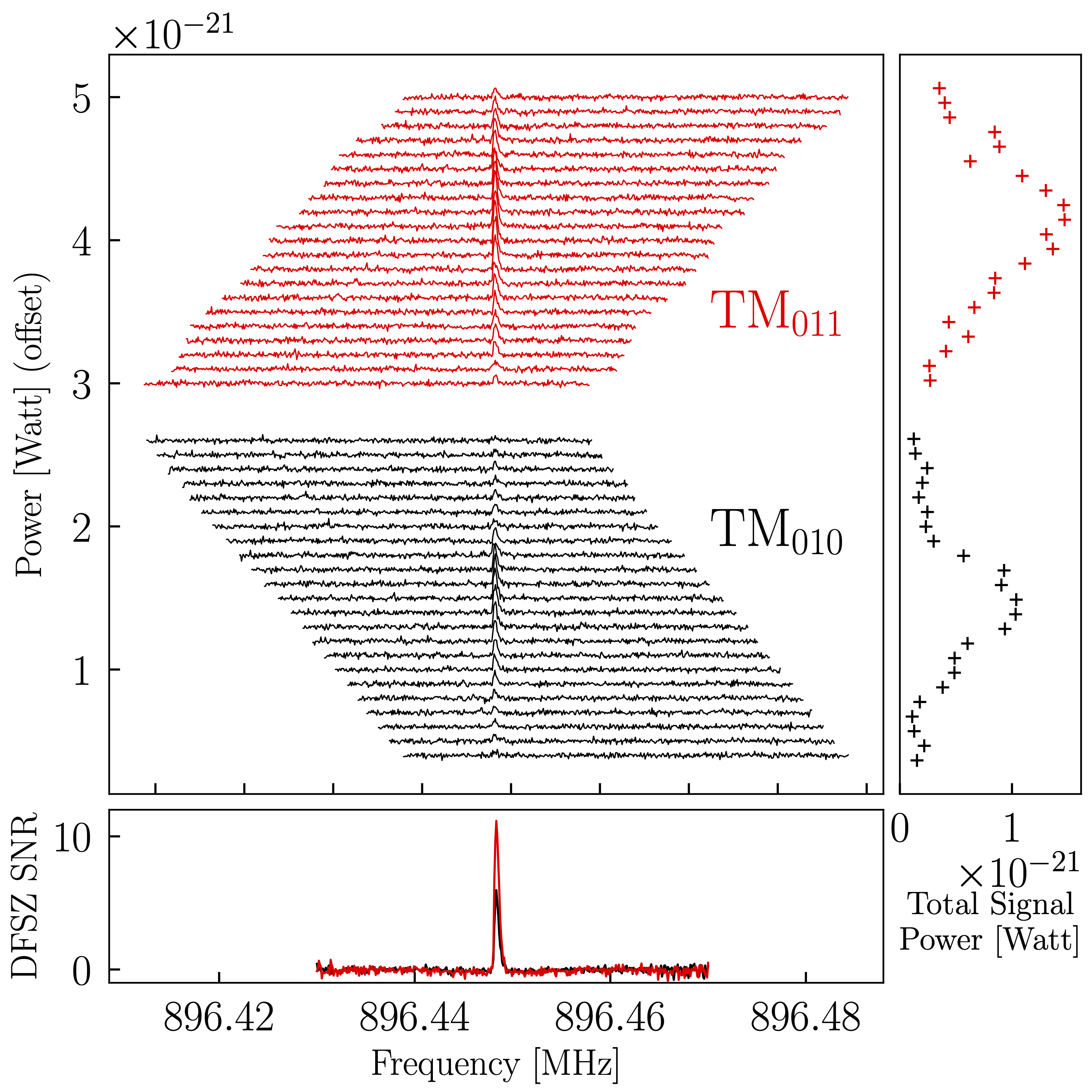}
    \caption{Digitization spectra including a blinded SAG for $\rm TM_{010}$ (black) and $\rm TM_{011}$ (red) mode after removing the receiver shape and JPA standing wave distortion. The bottom small plot shows combined SNR with respect  to DFSZ power. The right small plot shows the integrated signal power and how it varies according to the Lorentzian cavity enhancement.}
    \label{fig:SNRoverlay}
\end{figure}

A single candidate precipitated the full series of studies that is undertaken in the event of an axion-like signal, for which the final step would be a magnet ramp to check for $B^2$ scaling of the candidate signal power, as expected for an axion signal. Figure~\ref{fig:SNRoverlay} shows the lineshapes for the candidate. The candidate consistently emerged at the same frequency even though the cavity resonant frequency and digitization window were shifted. The frequency integrated power followed the expected Lorentzian lineshape, consistent with the quality factor of the cavity. Therefore, it most likely originates inside the cavity. The individual lineshapes followed the Maxwell-Boltzmann distribution exactly within the statistical uncertainty, as is expected from the standard halo model for dark matter. 

\begin{figure}[phtb]
    \centering
    \includegraphics[width=\linewidth]{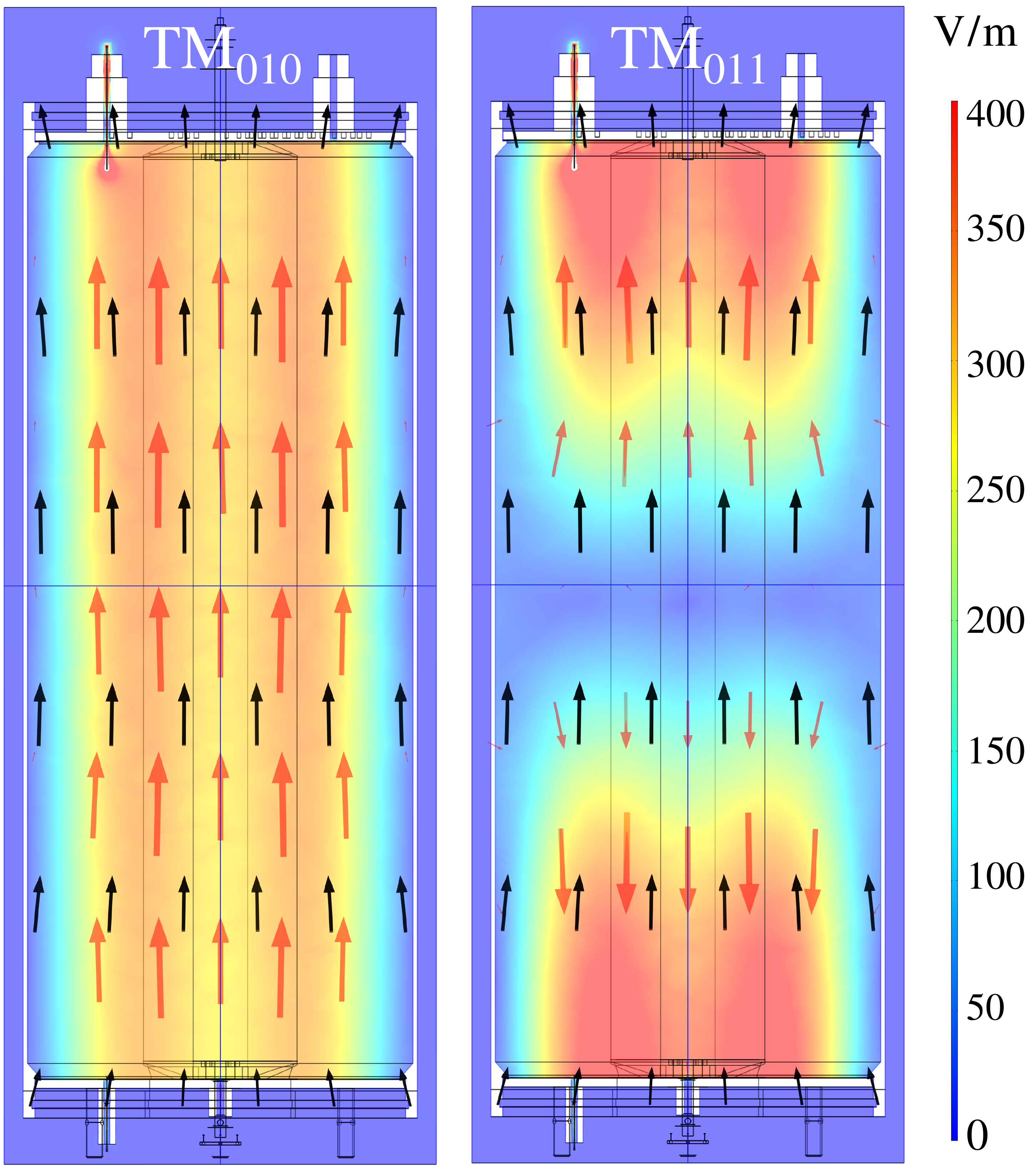}
    \caption{Electric field distribution as a heat map for $\rm TM_{010}$ (left) and $\rm TM_{011}$ (right) modes simulated with COMSOL Multiphysics~\cite{comsol}. Electric field for the mode and impressed magnetic field are shown as red and black arrows, respectively. Calculated form factors with CST Magnetic Field Solver are 0.455 and $<0.001$ for $\rm TM_{010}$ and $\rm TM_{011}$ modes, respectively. The overlap between the $\rm TM_{010}$ electric field and the magnetic field is large and consistent across the volume, while the overlap between the $\rm TM_{011}$ electric field is of opposite sign in the top and bottom of the cavity, leading to cancellations in the cavity response to the spatially uniform dark matter axion field.}
    \label{fig:TM010TM011}
\end{figure}

A true axion signal should disappear when the $\rm TM_{011}$ mode is tuned to the candidate frequency because the electric fields are of the opposite polarity at the top and bottom of the cavity as shown in Fig.~\ref{fig:TM010TM011}, so the form factor is almost zero. The candidate failed this test, and the magnet ramp was not applied. Therefore, the candidate was determined not to be consistent with the axion hypothesis and was subsequently revealed to be a blinded SAG. 

\begin{table}[phtb]
\caption{\label{tab:summary_candidates} A list of candidates remaining after turning off calibration SAGs. The 896.448-MHz candidate was a blinded SAG. ``Persistence'' is checked when the candidate exists in all the scans with similar powers. ``At Same Frequency'' is checked when the candidate was at the given frequency $\pm 300$~Hz. ``Not in Air'' is flagged if the candidate could not be observed with a spectrum analyzer attached to an external antenna at the experimental site. ``Enhanced on Resonance'' is flagged when the integral of the signal power was scaled as a Lorentzian function. ``$\times$" denotes tested but not passed.}
\begin{ruledtabular}
\begin{tabular}{lcccc}
    \shortstack[c]{Frequency\\ $\rm [MHz]$} & \shortstack[c]{Persistence} &  \shortstack[c]{At Same\\ Frequency} &\shortstack[c]{Not\\ in Air} &  \shortstack[c]{Enhanced\\ on Resonance}  \\
    \colrule
     839.669 & \Checkmark &  $\times$  & \Checkmark &  $\times$  \\
     840.268 & \Checkmark & \Checkmark & \Checkmark &  $\times$  \\
     860.000 & \Checkmark & \Checkmark &  $\times$  &  $\times$  \\
     891.070 & \Checkmark & \Checkmark & \Checkmark &  $\times$  \\
     896.448 & \Checkmark & \Checkmark & \Checkmark & \Checkmark\\
     974.989 &  $\times$  & \Checkmark & \Checkmark &  $\times$  \\
     974.999 &  $\times$  & \Checkmark & \Checkmark &  $\times$  \\
     960.000 & \Checkmark & \Checkmark &  $\times$  &  $\times$  \\
     980.000 & \Checkmark & \Checkmark &  $\times$  &  $\times$  \\
     990.000 & \Checkmark & \Checkmark &  $\times$  &  $\times$  \\
     990.031 &  $\times$  & \Checkmark & \Checkmark &  $\times$  \\
    1000.000 & \Checkmark & \Checkmark &  $\times$  &  $\times$  \\
    1000.013 &  $\times$  & \Checkmark & \Checkmark &  $\times$  \\
    1010.000 & \Checkmark & \Checkmark &  $\times$  &  $\times$  \\
    1020.000 & \Checkmark & \Checkmark &  $\times$  &  $\times$  \\
\end{tabular}
\end{ruledtabular}
\end{table}

Other than the blinded SAG, the analysis procedure detected 15 persistent candidates as summarized in Table~\ref{tab:summary_candidates}. None of the candidates passed the ``Enhanced on Resonance'' check except the blinded SAG described above. The other candidates did not pass the other criteria for being likely axion candidates. Therefore we are confident that these signals did not result from axion dark matter.

\begin{figure}[phtb]
    \centering
    \includegraphics[width=\linewidth]{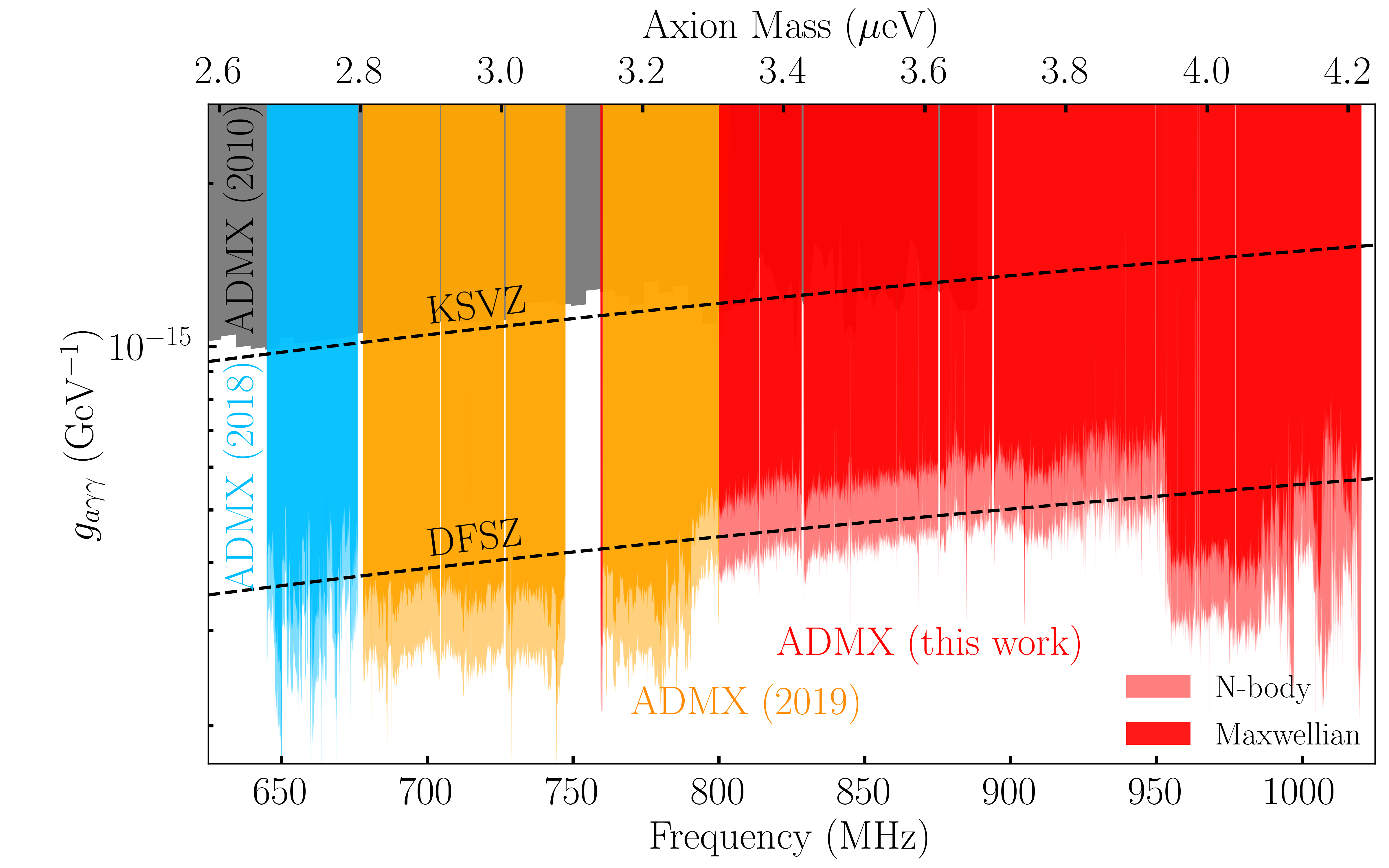}
    \caption{90\% C.L. upper limits on \gagamgam as a function of axion mass. The gray-, blue-, and yellow-colored areas represent previous ADMX limits reported in Ref.~\cite{PhysRevD.64.092003}, \cite{PhysRevLett.120.151301}, and \cite{PhysRevLett.124.101303}. The red-colored area shows the limits of this work. We ruled out KSVZ (DFSZ) axions in the $3.3\text{--}4.2~(3.9\text{--}4.1)\si\micro$eV mass range.}
    \label{fig:limit}
\end{figure}

Upon eliminating all candidates in the frequency range of interest, we set 90\% confidence level (C.L.) upper limits on the axion-photon couplings across the explored mass ranges with the assumption that axions make up 100\% of the local dark matter density as shown in Fig.~\ref{fig:limit}. We ruled out KSVZ (DFSZ) axions in the $3.3\text{--}4.2~(3.9\text{--}4.1)~\si\micro$eV mass range. We initially aimed for DFSZ sensitivity overall (starting near the high-mass end of the energy range), but our system noise was suboptimal, making the scan speed much slower than we would like. Consequently, we decided to cover the remaining frequency range at 2 times DFSZ sensitivity (by scanning much more quickly) and then to upgrade the detector to reduce the system noise. The limits assumed that the velocity distribution of dark matter axions is either a boosted Maxwell-Boltzmann distribution~\cite{PhysRevD.42.3572} with dark matter density $\rm 0.45~GeV/cc$ or in accordance with an N-body simulation~\cite{0004-637X-845-2-121} with dark matter density $\rm 0.6~GeV/cc$. The limits include systematic experimental uncertainties associated with the cavity, the amplifiers, and the electromagnetic field simulation as shown in Table~\ref{tab:uncer_summary}. The largest uncertainty was \ToffEp dominated by the temperature sensor accuracy. Additionally, a sensitivity loss from potential over-fitting of signals from the Pad\'{e}-approximant used to remove distortions in the digitized spectra was taken into account. This effect was quantified by injecting software synthetic signals into the real data and comparing analysis results to the injected power. We determined that there was a 20\% suppression of the axion power, i.e. $\sim 10\%$ loss of the axion-photon coupling.

\begin{table}[phtb]
\caption{\label{tab:uncer_summary} Summary of uncertainties associated with the observed power.}
\begin{ruledtabular}
\centering
\begin{tabular}{lc}
    Source & Fractional Uncertainty on \Paxion\\
   \colrule
   Cavity Q-factor &  2 \%\\
   Antenna Coupling & 2 \% \\
   JPA SNRI & ~~0.8 \%\\
   \ToffEp & ~~4.3 \%\\
   $B^2 VC$ & 3 \% \\
   Total & 6 \% \\
\end{tabular}
\end{ruledtabular}
\end{table}

\begin{figure}[phtb]
    \centering
    \includegraphics[width=\linewidth]{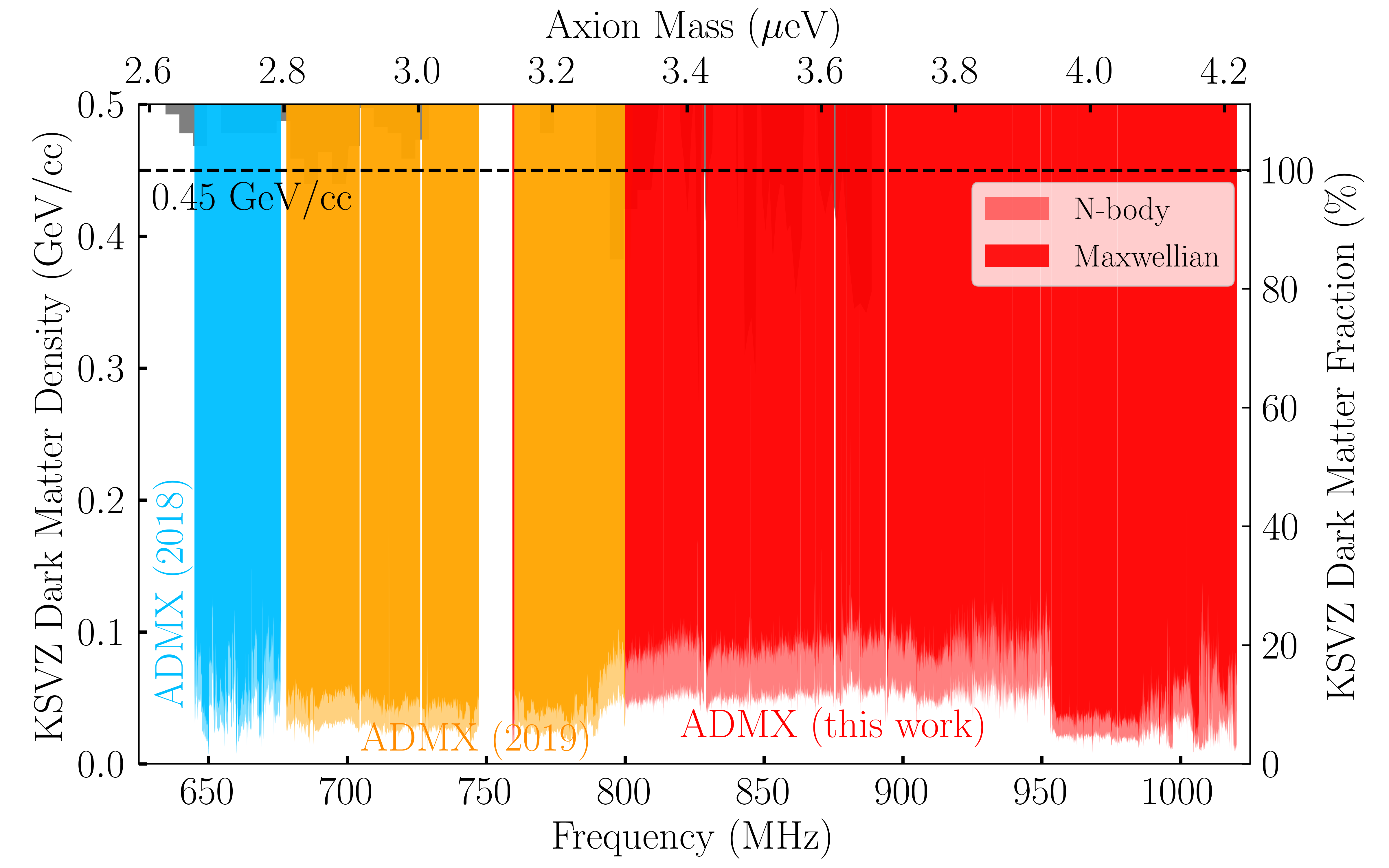}
    \caption{90\% C.L. upper limits on the dark matter energy density assuming the KSVZ model for axion coupling. The blue- and yellow-colored areas represent previous ADMX limits reported in Ref.~\cite{PhysRevLett.120.151301} and \cite{PhysRevLett.124.101303}, respectively. The red-colored area shows the limits of this work. The KSVZ axions are excluded even though the axion density is 0.1~GeV/cc (20\%) of the total dark matter density.}
    \label{fig:limit_rho}
\end{figure}

Alternatively, we can set limits on the dark matter axion density with the assumption that the axion-photon coupling is given by the KSVZ model. This is shown in Fig.~\ref{fig:limit_rho}. The limit shows that KSVZ axions are excluded from contributing any more than $\rm 0.1~GeV/cc$ of the local dark matter density, or 20\% of the expected density.

In summary, we searched for the ``invisible" axion dark matter in the \runOneClimitMass mass range. No axion-like excess was observed. Therefore, we set a limit for the axion-photon coupling which is the most stringent to date. We intend to rescan the region covered here (\runOneClimitMass) at DFSZ sensitivity after making a number of upgrades and repairs to the current cavity and RF system. These upgrades will include improving the thermal isolation of the cavity from the 1-K support to lower the overall heat load on the dilution refrigerator, upgrading the $\mu$-metal shield to mitigate the remaining magnetic field on the JPA, and adding additional temperature sensors to measure the system noise temperature precisely. 

This work was supported by the U.S. Department of Energy through Grants No. DE-SC0009800, No. DESC0009723, No. DE-SC0010296, No. DE-SC0010280, No. DE-SC0011665, No. DEFG02-97ER41029, No. DEFG02-96ER40956, No. DEAC52-07NA27344, No. DEC03-76SF00098, and No. DE-SC0017987. Fermilab is a U.S. Department of Energy, Office of Science, HEP User Facility. Fermilab is managed by Fermi Research Alliance, LLC (FRA), acting under Contract No. DE-AC02-07CH11359. Additional support was provided by the Heising-Simons Foundation and by the Lawrence Livermore National Laboratory and Pacific Northwest National Laboratory LDRD offices. UWA participation is funded by the ARC Centre of Excellence for Engineered Quantum Systems, CE170100009, Dark Matter Particle Physics, CE200100008, and Forrest Research Foundation. The corresponding author is supported by JSPS Overseas Research Fellowships No. 202060305. LLNL Release No. LLNL-JRNL-826807. 
\newpage

\bibliographystyle{apsrev4-2}
\bibliography{references}

\end{document}